\documentclass[12pt]{article}
\providecommand{\href}[2]{#2}   
\newcommand{\be}[3]{\begin{equation}  \label{#1#2#3}}

\newcommand{\ee}{ \end{equation}}
\newcommand{\ba}{\begin{array}}
\newcommand{\ea}{\end{array}}

\renewcommand{\arraystretch}{1.7}
\begin{document}
\begin{flushright}
HUB-EP-98/77 \\
hep-th/9812169  
\end{flushright}
\vspace{1cm}

\centerline{\bf \Large
Orbifolds of $AdS_3$ and fixpoints of the CFT}


\vspace{1cm}

\centerline{\bf Klaus Behrndt\footnote{Email: behrndt@physik.hu-berlin.de ; 
    Talks presented at the ``32nd International
    Symposium Ahrenshoop on the Theory of Elementary Particles'', Buckow
    1998 and at "Quantum aspects of gauge theories,
    supergravity and unification", Corfu, September 1998.
Work is supported by the DFG.}
}


\bigskip

\centerline{\em Humboldt-University Berlin, Invalidenstrasse 110,
10115 Berlin, Germany}


\vspace{1cm}

\noindent
\begin{abstract}
The 3-d BTZ black hole represents an orbifold of $AdS_3$ gravity.
The UV as well as the IR region of the CFT is governed by a
gauged $SL(2, {\bf R})$ WZW model. In the UV it corresponds to
a light-cone gauging (Liouville model) whereas in the IR it is 
a space-like gauging (2-d black hole).
\end{abstract}

\vspace{5mm}

%
%
\section{Introduction and summary}


In the AdS/CFT correspondence \cite{050}, \cite{062}, \cite{390} the
radial coordinate of the AdS space sets the energy scale of the world
volume field theory.  The brane does not reside at infinity, instead
the complete AdS space is expected to be dual (in proper limits) to
the world volume field theory \cite{053}. In the maximal
supersymmetric case the world volume field theory is scale invariant
and thus the field theory is conformal for any energy scale.  This
means in the AdS picture that we could have set a cutoff at any other
radius, it would reproduce the same boundary conformal field theory
(CFT).

The situation changes however if we break supersymmetry, which can be
done by orbifolding either in the spherical part or in the AdS part.
Consider e.g.\ the case of $AdS_3 \times S_3$.  As we will discuss
below AdS orbifolding introduces a black hole, which obviously break
the scale invariance.  To be concrete, the orbifold $AdS_3/Z_n$
describes a BTZ black hole \cite{160} (see also \cite{410}) and it
corresponds to the near-horizon geometry of a black string with
momentum modes, where $n$ is the momentum number. On the other hand an
$S_3$ orbifold corresponds to the embedding of a Taub-NUT space, which
changes also the AdS part. E.g.\ the $AdS_3$ part of the $D1 \times
D5$ intersection reads $ds^2 = r^2 du dv + (dr/r)^2$ and it becomes
$ds^2 = r du dv + (dr/r)^2$ for $D1 \times D5$+Taub-NUT, i.e.\ the
power of $r$ has changed. This is obvious due to the different
harmonic functions.  So, this orbifolding breaks
the scale invariance $r \rightarrow e^{\lambda_D} r$ and $x_{i}
\rightarrow e^{-\lambda_D} x_{i}$, where $x_i$ are the world volume
coordinates. Of course, in both cases we can restore the scale
invariance locally by a proper coordinate transformation, but
nevertheless they are globally different. Notice, coordinate
transformations in the AdS space corresponds to an operator
reparameterization of the boundary field theory.

In this letter we show, that the boundary CFT in the UV as well as in
IR can be described by different gauged WZW models. In AdS gravity the
asymptotic region (large radius) translates into the UV region of the
CFT and the IR corresponds to the near-horizon region of the BTZ black
hole.  The $AdS_3$ orbifolding fixes the subgroup, that has to be
gauged: in the UV (asymptotic AdS region) the CFT is given by an
${SL(2, {\bf R}) \over SO(1,1)}$ WZW model and in the IR (near horizon
region) one gets an ${SL(2, {\bf R}) \over U(1)}$ WZW model. Since the
central charges of both CFT's differ, there is no exactly marginal
deformation connecting both regions, i.e.\ there is a non-trivial
renormalization group flow.  At many points we will be very brief and
refer to \cite{380} and \cite{020} for more details and more
references.


\section{BTZ black hole as $AdS_3$ orbifolding}


A three-dimensional anti-de Sitter space-time is defined as a
hyperboloid in a 4-d space with the signature $(-++-)$, i.e.
\be010
-l^2 = -(X^0)^2 + (X^1)^2 + (X^2)^2 - (X^3)^2 = - z_1^+ z_1^- + z_2^+ z_2^-
\ee
where we introduced the coordinates 
\be020
z_1^{\pm} = l  \, e^{\pm {1 \over 2} (\theta_R
 + \theta_L)} \, \cosh {\lambda \over 2} \qquad , \qquad
z_2^{\pm} = l \, e^{\pm {1 \over 2} (\theta_R
 - \theta_L)}\, \sinh {\lambda \over 2}  \ .
\ee
The metric reads
\be030
ds^2 = - dz_1^+ dz_1^- + dz_2^+ dz_2^- = {l^2 \over 4} 
\Big(d\lambda^2 + d\theta_R^2 + d \theta_L^2 + 2 \cosh\lambda \, d\theta_R
d\theta_L \Big) \ .
\ee
Defining $\theta_R = (y + t)/l$, $\theta_L = ( y -t)/l$ and $\lambda = 2 \rho$
the metric can also be written as
\be040
ds^2 = l^2 d\rho^2 + \cosh^2\rho \, dy^2 - \sinh^2\rho \, dt^2  \ .
\ee
Furthermore, the $AdS_3$ space is the $SL(2, {\rm \bf R})$ group
space and a group element $g$ can be written as
\renewcommand{\arraystretch}{1.2}
\be050
g =     e^{\theta_L T_1} \; e^{ \lambda T_2} \; e^{ \theta_R T_1} \ = \
        \left(\ba{cc} \cosh{\theta_L \over 2} & \sinh{\theta_L\over 2} \\
        \sinh{\theta_L \over 2} & \cosh{\theta_L \over 2} \ea \right)
        \left(\ba{cc} e^{\lambda/2} & 0 \\ 0 & e^{- \lambda/2} \ea \right)
        \left(\ba{cc} \cosh{\theta_R \over 2} & \sinh{\theta_R \over 2} \\
        \sinh{\theta_R \over 2} & \cosh{\theta_R \over 2} \ea \right)
\ee
\renewcommand{\arraystretch}{1.7}
with the $SL(2, \bf R)$ algebra
\be060
 [ T_a , T_b ] = \epsilon_{ab}^{ \ \  c} T_c, \hspace{0,5cm}
 {\rm Tr} (T_a T_b) = \frac{1}{2} \eta_{ab}
\ee
($\eta= {\rm diag}(-1,1,1)$, $\epsilon^{012}=1$). A
representation is
\renewcommand{\arraystretch}{1}
\be070
  T_0 = \frac{1}{2} \ \left (
   \begin{array}{rr}
   0     &   -1   \\
   1     &   0  \\
\end{array}
\right), \ \ \ \
T_1 = \frac{1}{2} \
\left (
   \begin{array}{rr}
   0     & 1   \\
   1     & 0  \\
\end{array}
\right ), \ \ \ \
T_2 = \frac{1}{2} \
\left (
   \begin{array}{rr}
   1    & 0   \\
   0    & -1  \\
\end{array}
\right )\ .
\ee
\renewcommand{\arraystretch}{1.7}
In terms of the $SL(2)$ current
\be080
\ba{rcl}
J = J^a T_a = {l \over 2} \, g^{-1} dg &=& {l \over 2}
   \, [-\sinh \theta_L \; d\lambda + \sinh \lambda \,
        \cosh  \theta_L \; d\theta_R ] \, T_0  \\
  && + {l \over 2} \, [\cosh \lambda \; d\theta_R + d\theta_L ] \, T_1 \\
  && +  {l \over 2} \, [\cosh  \theta_L \; d\lambda - \sinh \lambda \,
    \sinh  \theta_L \; d\theta_R ] \, T_2
\ea
\ee
the $AdS_3$ metric (\ref{030}) becomes
\be090
ds^2 = tr(J^2) = - (J^0)^2 + (J^1)^2  + (J^3)^2 \ .
\ee
Let us now turn to the BTZ black hole \cite{160} which reads
\be100
ds^2 = - e^{-2V(r)} \ dt^2 + e^{2V(r)} \ dr^2 + \Big({r\over l}\Big)^2 \
        \Big( dy -\frac{r_- r_+}{ r^2} \, dt  \Big)^2
\ee
with
\be110
e^{-2 V(r)} = {(r^2 - r_-^2)(r^2 - r_+^2) \over r^2 l^2} \ .
\ee
This black hole can also be written in terms of two chiral currents
\be120
\begin{array}{l}
{l \over 2} \, g^{-1} d g \equiv {l \over 2} \, A = {1 \over 2}
\Big( e^{-V} \, T_0  + {r \over l} (1 - {r_- r_+ \over r^2}) \, T_1 \Big)
        d v + {1 \over 2} e^{V} (1 + {r_- r_+ \over r^2})\, T_2  \
        dr \ , \\
{l \over 2} \, \bar g^{-1} d \bar g \equiv {l \over 2}
\bar A = {1 \over 2} \Big( e^{-V} \, T_0  - {r \over l} 
	(1 + {r_- r_+ \over r^2}) \,
        T_1 \Big) du   - {1 \over 2} e^{V} (1 - {r_- r_+ \over r^2}) \,
        T_2 dr \ .
\end{array}
\ee
with $g, \bar g \in SL(2, {\bf R})$ and 
\be122
v/u = l \, \theta_{R/L} = y \pm t \ .
\ee
Combining the chiral currents to a non-chiral, the metric of
the BTZ black hole is
\be130
ds^2 = {l \over 2} tr( A - \bar A)^2 \ .
\ee
For the BTZ black hole the $y$ coordinate is periodic, which can be
expressed by $y \simeq y + {4 \pi l \over n}$. This means
\be140
\theta_{R/L} \simeq \theta_{R/L} + {4 \pi \over n}
\qquad {\rm or} \qquad z_1^{\pm} \simeq e^{\pm {4 \pi \over n}} z_1^{\pm}
\ee
which corresponds to an orbifolding for the $AdS_3$ space. Notice, 
this orbifold breaks all supersymmetries, which coincides with
the non-extremality of the BTZ black hole metric (\ref{100}).  In the
next section we will see, that in the supersymmetric case ($r_+ =
r_-$) one chiral current drops out and we get an orbifolding only in
$\theta_L$
\be142
\theta_{L} \simeq \theta_{L} + {4 \pi \over n}
\qquad {\rm or} \qquad z_1^{\pm} \simeq e^{\pm  {2 \pi \over n}} z_1^{\pm}
\quad , \quad z_2^{\pm} \simeq e^{\mp  {2 \pi \over n}} z_2^{\pm} \ .
\ee
This orbifold can be seen as lens space for $AdS_3$, by
appropriate Wick rotation we can obtain the $S_3$ lens space
\cite{420}.


\section{Conformal fixpoints}


In the previous section we have argued that ``adding'' a BTZ black
hole to the $AdS_3$ space corresponds to an orbifolding. One may
regard this black hole as a (small) perturbation of the asymptotic CFT
and the corresponding states, which are invariant under this orbifold,
can be related to microstates of the BTZ black hole and therewith to
4-d black holes, see e.g.\ \cite{360}, \cite{370}.  On the other hand,
since the radial coordinate corresponds to the energy scale, this
asymptotic CFT describes the UV fixpoint.  One goal of this letter is,
to investigate the IR fixpoint and the CFT that describes this
fixpoint. Obviously, a flow towards the IR corresponds to the radial
movement and if there is a further fixpoint we may expect them near
the horizon.

To make this more explicit we can employ the fact, that 3-d AdS
gravity can be described by a Chern-Simons theory and is exactly
solvable \cite{170}, \cite{180}.  Decomposing the $AdS_3$
diffeomorphism group $SO(2,2) \simeq SL(2,{\bf R})_L \times SL(2, {\bf
R})_R$, the 3-dimensional action can be written as
\be150
S = S_{CS} [A] \ - \  S_{CS} [\bar A]
\ee
with
\be160
S_{CS} [A] = \frac{k}{4 \pi} \int_{M_3} d^3 x
        {\rm Tr} \ \Big(AdA + \frac{2}{3} A^3 \Big) 
\ee
where we will adopt the notation where the level $k = l$ (in other
notations $k= 2\pi l/ \kappa_3$).  The gauge
field one-forms are
\be170
A = (\omega^a + \frac{1}{l} e^a) \ T_a \ \in SL(2,{\bf R})_R \ , \hspace{1cm}
\bar A = ( \omega^a - \frac{1}{l} e^a) \ \bar T_a \ \in SL(2,{\bf R})_L.
\ee
where $\omega^a \equiv \frac{1}{2} \epsilon^{abc} \omega_{bc}$ are
given by the spin-connections $\omega_{bc}$ and $e^a$ are the
dreibeine. Due to boundaries the Chern-Simons theory is not invariant
under gauge transformations and as a consequence gauge degrees of
freedom do not decouple and become dynamical on the boundaries. These
are the degrees of freedom of the conformal field theories living at
the boundaries.

In the following we will discuss this procedure for the BTZ black hole.
The geometry of the manifold is $M_3 = {\bf R} \times \Sigma$, where
${\bf R}$ corresponds to the time of the covering space of $AdS_3$ and
$\Sigma$ represents an ``annulus'' $r_+ \leq r < \infty$.

Calculating the gauge connections $A= A^a T_a$ and $\bar A = \bar A^a
\bar T_a$ for the BTZ solution (\ref{100}) one finds exactly the
fields (\ref{120}), for details see \cite{020}.  In order to extract
the CFTs at the boundaries we have to perform two steps: (i) we have
to add boundary terms that impose the correct boundary conditions and
(ii) we have to mod out the isometry group related to the orbifold, that
corresponds to the Killing direction that has been periodically
identified in constructing the BTZ black hole.

(i) As dictated by the Chern-Simons solution (\ref{120}) we will impose
as boundary conditions
\be180
A_u = \bar A_v =0
\ee
and therefore we add as boundary term to the action
\be190
\delta S = {k \over 8\pi} \int_{\partial M} Tr
(A_v A_u + \bar A_v \bar A_u) \ .
\ee
Since we have flat gauge connection we insert $A = g^{-1} dg$ and
$\bar A = \bar g^{-1} d \bar g$ into the action and obtain as
result an $SL(2, {\bf R})$ WZW model \cite{150}. 

(ii) In a second step we translate the orbifolding (\ref{140}) into a
gauging of an $SL(2, {\bf R})$ group direction. {From} the
Chern-Simons fields (\ref{120}) we find
\be200
\ba{rl}
{\rm for}\ r \rightarrow \infty:\quad &
  A \ = \ {r\over l} (T_1 + T_0)
  \; {dv \over l} + T_2 \; {dr \over r} \ = \ {r\over l} \, T_+
  \, {dv \over l} + T_2 \; {dr \over r}
  \\
  & \bar A \ =\ - {r\over l} (T_1 - T_0) \; {du \over l} - T_2 \; {dr \over r}
  \ = \ - {r\over l} \, T_- \, {du \over l} - T_2 \; {dr \over r} \ .
\ea
\ee
with $T_{\pm} = T_1 \pm T_0$.  Therefore, the orbifolding in
$\theta_{L/R} \sim u/v$ corresponds to a gauging of the lightcone
group directions $T_{\pm}$. On the other hand, in the IR ($\lambda
\rightarrow 0$) we reach the horizon boundary $r\rightarrow r_+$ and
the gauge fields become
\be210
\ba{ll}
{\rm for } \quad r \rightarrow r_+  \quad ({\rm or} \quad 
\lambda \rightarrow 0)  :
 \quad &
  A \ = \ {1 \over l} (r_+ - r_-) \; T_1  \; {dv \over l} +
  \; T_2 \; d\lambda \\
  & \bar A \ =\ - {1\over l}(r_+ + r_-) \; T_1 \; {du \over l} -
  \; T_2 \; d\lambda 
\ea
\ee
where $\lambda$ is defined by the radial coordinate for which $A_r$
and $\bar A_r$ are constant. So we see, that in this case the
orbifolding translates into a gauging of the $T_1$ group direction.

Both gauged WZW models have been discussed some time ago.  The gauging
of a lightcone group direction, i.e.\ an ${SL(2, {\bf R}) \over
SO(1,1)}$ WZW model corresponds to a Liouville model \cite{400},
\cite{210}, where the Liouville field corresponds to the radial
coordinate \cite{211}; see also \cite{390}.  The gauging of a spatial
direction yields an ${SL(2, {\bf R}) \over U(1)}$ WZW model, which
describes a 2-d black hole solution \cite{140}, \cite{210}. Both
solutions are known to be exact CFTs, but with different central
charges.  For the Liouville model, describing the CFT in the UV
region, the central charge is \cite{210}
\be220
c_{UV} = 1 + 6(k-2) Q^2 = {3k \over k-2} - 2 + 6k 
\ee
where $Q={k -1 \over k - 2}$ is the background charge of the Liouville field
($k = {l \over 4 \alpha'}$).
On the other hand the CFT in the IR has the central charge
\be260
c_{IR} = {3k \over k-2} - 1 \ .
\ee
Due to the difference in the central charges there cannot be an
exactly marginal deformation between both CFT's. So, there is a
non-trivial renormalization group flow (see also \cite{250}), which
corresponds to the bulk physics of the BTZ black hole.  Due to the
holographic nature, the complete bulk physics will be fixed by these
boundary CFTs.  However, let us stress that at any finite point in
space time one can promote the background to an exact CFT. On one hand
the gauged WZW model can be made exact by changing the renormalization
group scheme (field redefinitions) \cite{320} and on the
other hand the BTZ black hole is locally at any point $AdS_3$.
The different central charges indicate the non-trivial
global structure of the model and it is better thought of as an
interpolating solution between two (different) CFTs.

Finally, we will comment on the supersymmetric case where the BTZ
black hole becomes extremal, i.e.\ $r_- = r_+$.  In this case the
Chern-Simons field $A$ decouples and the CFT is coverned by one chiral
current ($\bar A$) only. In fact, if we first discuss the IR limit ($r
\rightarrow r_+$) we see from (\ref{210}) that $A_v = 0$ and due to the
boundary condition (\ref{180}) also $A_u =0$ and thus $A$ drops
completely ($A_r$ was fixed due to the choice of the coordinates).  On
the other hand in the UV we get the same asymptotic behavior as for
the non-extremal case. But by performing the orbifolding in $\theta_L$
(gauging of the $T_-$ direction) one truncates the theory already to
Liouville\footnote{By gauging a light-cone direction one gets rid of
two degrees of freedom \cite{210}.}, i.e.\ one has to gauge only one
light-cone direction.  Hence, in the extremal case the 
Chern-Simons field $A$ decouples
from the CFT in the IR as well as in the UV. These two points are
boundaries for the Chern-Simons theory and since gauge degrees of
freedom can become dynamical only on these boundaries, we can
disregard the $A$ field completely. Therefore, as expected the
supersymmetric case is governed by only one chiral Chern-Simons field
and corresponds only to the orbifold of $\theta_L$ in (\ref{142}).

\vspace{1cm}


%
%

\providecommand{\href}[2]{#2}\begingroup\raggedright\endgroup

\end{document}